\documentclass{article}
\usepackage{spconf,amsmath,graphicx,hyperref}
\usepackage{verbatim}
\usepackage{booktabs}
\usepackage{amssymb}

\title{A Cross-Lingual Acoustic Disease-Alignment Framework for Respiratory Health Assessment from Spontaneous Speech}
\name{\begin{tabular}{@{}c@{}}
Roksana Khanom$^{1}$,
Raghib Asfak Tasnim$^{2}$,
Bodrun Nahar Bithi$^{2}$,
Shafia Shirin Supty$^{3}$,\\
Saiful Islam Raju$^{4}$,
Ashok Agrawala$^{1}$,
Nirupam Roy$^{1}$
\end{tabular}}
\address{
$^{1}$University of Maryland, College Park, USA\\
$^{2}$Sylhet MAG Osmani Medical College Hospital, Bangladesh \\
$^{3}$Dr. M R Khan Shishu Hospital \& Institute of Child Health, Bangladesh \\
$^{4}$Line Reflection Ltd., Bangladesh
}
\begin{document}
\ninept
\maketitle
\begin{abstract}
Spontaneous speech offers a scalable, noninvasive signal for respiratory health assessment, yet interpretable models that generalize across languages remain challenging because disease-related acoustic changes are confounded by language-specific phonetic variation. We present CL-DAF, a Cross-Lingual Disease-Alignment Framework that identifies acoustic dimensions whose disease effects remain consistent across languages. Using 201 English and 75 newly collected Bangla speakers, we construct a common 272-dimensional acoustic representation and quantify disease alignment using signed rank-biserial effects and the Language Invariance Score. We first show that spontaneous Bangla speech separates COPD from controls (AUC 0.85); however, 133 features reverse their disease direction across languages and the full representation transfers poorly (AUC 0.49 from Bangla to English). CL-DAF isolates 26 disease-aligned features that raise AUCs to 0.825 and 0.722 from English to Bangla and Bangla to English, respectively. These findings provide a foundation for multilingual clinical speech models emphasizing pathology over language-dependent variation.
\end{abstract}

\begin{keywords}
cross-lingual speech, respiratory health, COPD, acoustic biomarkers,
multilingual clinical AI
\end{keywords}
\begin{figure*}[t]
    \centering
    \includegraphics[
        width=2.06\columnwidth,
        keepaspectratio
    ]{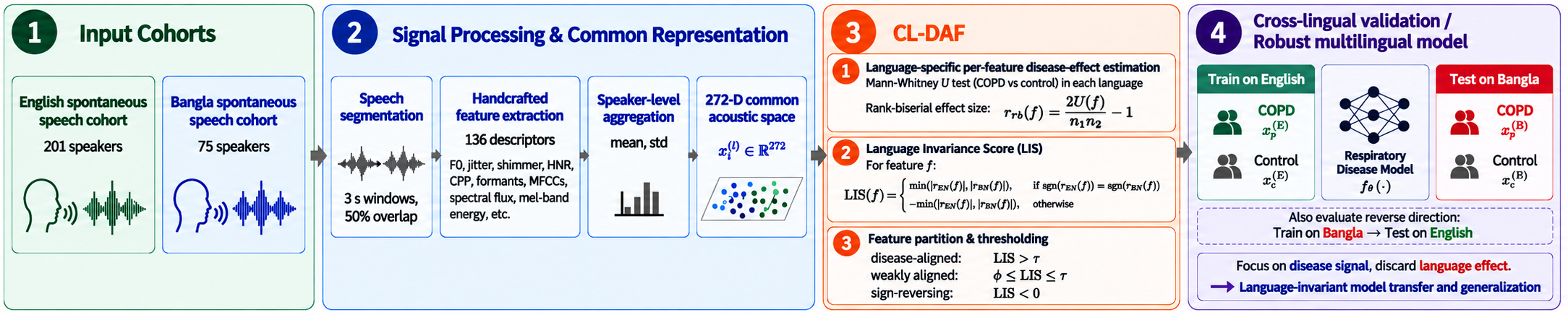}
    \vspace{-0.1in}
    \caption{Overview of the proposed CL-DAF framework for cross-lingual COPD screening.}
    \label{fig:framework}
    \vspace{-0.1in}
\end{figure*}
\section{Introduction}
\label{sec:intro}

The human voice is a promising non-invasive biomarker because
speech production depends on the coordinated function of the
respiratory, phonatory, and articulatory systems~\cite{fuchs2021respiratory}.
This connection is particularly relevant to chronic obstructive
pulmonary disease (COPD), where persistent airflow limitation can
alter the respiratory support required for phonation and
speech~\cite{mayr2025assessing,nallanthighal2022detection}. Beyond linguistic content, acoustic signals can encode physical and contextual information, motivating their use for physiological sensing in respiratory conditions such as COPD \cite{nallanthighal2022detection}. This disease remains the third leading cause
of death worldwide and is frequently diagnosed only after disease
progression~\cite{who2026copd}, motivating
low-burden approaches for earlier assessment and longitudinal
monitoring. Prior studies have shown that respiratory impairment
can be detected from sustained vowels~\cite{idrisoglu2024copdvd,
triantafyllopoulos2024sustained}, cough and breathing
sounds~\cite{haider2019respiratory}, and other
controlled vocal tasks, while recent work has
demonstrated that natural speech also contains measurable
information related to respiratory function~\cite{khanom2026spirophonia,bhalla2025phoneme}. In particular, SpiroPhonia~\cite{khanom2026spirophonia} demonstrated that
unconstrained English spontaneous speech contains discriminative
COPD-related acoustic, spectral, and temporal information.

However, most respiratory voice studies have been developed
within single-language cohorts~\cite{idrisoglu2024copdvd,triantafyllopoulos2024sustained}, leaving the applicability
of spontaneous-speech respiratory biomarkers to linguistically
diverse populations largely unexplored. Bangla (also known as Bengali), spoken by over 270 million people worldwide and among the ten most widely spoken languages, represents both a linguistically and clinically important test case \cite{ethnologue2026}. Its phonological system includes systematic contrasts in voicing and aspiration across multiple places of articulation, providing an acoustically distinct setting from English for evaluating disease-related speech biomarkers \cite{ud2010bengali}. The clinical need is substantial, as COPD is highly prevalent and underdiagnosed in Bangladesh, with tobacco smoking and biomass-fuel exposure among its major modifiable risk factors~\cite{alam2015prevalence}. Because diagnosis depends on spirometry, which remains limited in availability, scalable speech-based assessment could provide a low-burden complement for repeated screening and longitudinal monitoring. Despite this linguistic and clinical relevance, 
no prior study has systematically investigated
COPD-related biomarkers in spontaneous Bangla speech. 

A broader challenge is that respiratory speech models are typically developed within individual languages, so extending assessment to a new language often requires a separately collected and clinically labeled cohort. Such datasets are costly and difficult to obtain, making independent language-specific models difficult to scale. Cross-lingual clinical-speech modeling has therefore begun to receive attention in other conditions, including Parkinson’s \cite{favaro2023multilingual} and Alzheimer’s diseases \cite{luz2023multilingual}, but comparable analysis for respiratory disease remains limited. This motivates multilingual respiratory models that can learn disease-related information in one language and generalize to another with limited target-language data. Clinical AI imposes a further requirement: beyond generalizing across populations, a model should indicate which acoustic changes drive its predictions~\cite{rudin2019stop}, a degree of transparency that self-supervised embeddings do not provide. An effective multilingual model should therefore preserve interpretable disease-related information while minimizing language-specific variation.

This problem is especially acute for spontaneous speech, where pathological changes are superimposed on language-dependent phonetic, prosodic, rhythmic, and temporal variation~\cite{bradlow2011language}. Consequently, an acoustic feature that
separates COPD from healthy speech in one language may become
weaker or even reverse its disease association in another; such
cross-lingual polarity reversals have recently been observed in
pathological speech representations~\cite{wang2026cross}. A model trained on such features may therefore follow a
language-dependent direction rather than a stable pathological
one. Thus, high within-language discriminability alone does not
establish that a feature is suitable for multilingual clinical
modeling. This raises a critical question: even when COPD is detectable within
each language independently, which components of that disease signal
remain reliable when the language changes? We address this question by first establishing that spontaneous
Bangla speech independently carries a detectable COPD signal, and
then formulating cross-lingual respiratory modeling as a
disease-alignment problem: \textit{which
acoustic dimensions preserve their COPD-related behavior across
languages, and which may hinder multilingual transfer?} The contributions of this paper can be summarized as follows:

\begin{enumerate}
\item We introduce
the first physician-labeled Bangla spontaneous-speech cohort for COPD assessment (75 speakers) and demonstrate that Bangla spontaneous speech alone carries discriminative COPD-related information, achieving a within-language AUC of 0.849. Paired with the English SpiroPhonia cohort~\cite{khanom2026spirophonia}, this yields a common 272-dimensional English–Bangla setting. 

\item We introduce CL-DAF (Fig.~\ref{fig:framework}), an interpretable framework that estimates signed disease effects independently within each language and uses the Language Invariance Score (LIS) to identify disease-aligned and direction-reversing features.

\item We characterize cross-linguistically stable respiratory biomarkers: CL-DAF selects 26 disease-aligned features, while 133 of the 272 features (48.9\%) reverse their COPD-versus-control direction between English and Bangla, demonstrating that {\em ``within-language discriminability does not necessarily imply cross-lingual disease consistency''}.

\item We validate disease alignment through bidirectional cross-lingual transfer. Despite strong within-language COPD separability, the full 272-feature representation transfers poorly, achieving AUCs of 0.663 for English$\rightarrow$Bangla and 0.488 for Bangla$\rightarrow$English. In contrast, the 26-feature CL-DAF representation improves AUCs to 0.825 and 0.722, respectively, while the sign-reversing subset degrades to 0.268 and 0.341. Target-held-out validation further supports generalization to unseen target speakers. 
\end{enumerate}

Together, these results show that multilingual clinical models should prioritize features whose pathological direction remains stable across languages rather than those selected solely for within-language predictability.

\section{Methods}
\subsection{Datasets and Cohort Construction} We study cross-lingual respiratory modeling using two independent
spontaneous-speech cohorts representing English and Bangla,
summarized in Table~\ref{tab:datasets}. The English domain is derived
from SpiroPhonia~\cite{khanom2026spirophonia}.
Since no suitable open Bangla-language respiratory spontaneous-speech
corpus was available, we collected a new clinical cohort in
collaboration with Sylhet MAG Osmani Medical College Hospital,
Bangladesh. Data were collected during participants'
hospital visits under an institutionally approved protocol
(IRB-2323776-2).
Respiratory labels were confirmed by treating physicians based on
clinical assessment and available medical records. The cohort
initially contained 76 participants; one control recording was
excluded after recording-level quality control, yielding 75 speakers
(26 COPD and 49 controls) for the final cross-lingual analysis.
\textbf{Dataset availability:} Available upon request for academic use.
\subsection{ Signal Processing and Common Acoustic Representation}
To construct a comparable feature space across the two linguistic
domains, all recordings were resampled to 16~kHz and processed using an
energy-based voice activity detector to retain speech regions. The
retained speech was divided into 3-s windows with 50\% overlap. From each
segment, we extracted 136 handcrafted descriptors using Parselmouth/Praat and Librosa~\cite{jadoul2018introducing,mcfee2015librosa}, spanning phonatory and
voice-quality measures (e.g., $F_0$, jitter, shimmer, HNR, CPP, and
formants) and spectral/cepstral characteristics (e.g., MFCCs and their
temporal derivatives, spectral centroid, bandwidth, roll-off, flux,
zero-crossing rate, and mel-band energies). Segment-level descriptors were aggregated to the speaker level using
their mean and standard deviation. For speaker $i$, the resulting
representation is

\begin{equation}
\mathbf{x}_i =
\left[
\boldsymbol{\mu}(\mathbf{z}_i),
\boldsymbol{\sigma}(\mathbf{z}_i)
\right]
\in \mathbb{R}^{272},
\end{equation}

where $\mathbf{z}_i$ denotes the 136-dimensional segment-level descriptor
set, and $\boldsymbol{\mu}(\cdot)$ and $\boldsymbol{\sigma}(\cdot)$ denote
the speaker-level mean and standard-deviation aggregation across segments.

Minimum/maximum and whole-recording temporal descriptors were excluded because recording durations differed
substantially between cohorts. Pause features were also excluded because non-participant speech in the Bangla recordings could confound pause statistics.
\vspace{-0.1in}
\begin{table}[t]
\centering
\caption{English and Bangla spontaneous-speech Cohort Summary.}
\label{tab:datasets}
\resizebox{\columnwidth}{!}{
\begin{tabular}{lcccccc}
\hline
\textbf{Dataset} &
\textbf{Disease} &
\textbf{Healthy} &
\textbf{Age (yr)} &
\textbf{Language} \\
\hline
SpiroPhonia
& 102
& 99
& 45--95
& English \\

Bangla Clinical
& 26
& 49
& 40--85
& Bangla \\
\hline
\end{tabular}}
\vspace{-0.2in}
\end{table}
\subsection{Cross-Lingual Problem Formulation}
\label{sec:problem_formulation}

Let $\mathcal{D}_{E}$ and $\mathcal{D}_{B}$ denote the English (EN) and
Bangla (BN) domains, respectively, represented in a common feature space. For each feature $f$, let
$\delta_{\ell}(f)$ denote its signed COPD--control effect in language
$\ell\in\{E,B\}$. A cross-lingual disease-effect reversal occurs when

\begin{equation}
\operatorname{sign}\!\left(\delta_E(f)\right)
\neq
\operatorname{sign}\!\left(\delta_B(f)\right),
\end{equation}

indicating that the same feature encodes opposing disease effects across
languages. CL-DAF therefore seeks a compact subspace that preserves both
disease-effect strength and pathological direction.

\subsection{CL-DAF: Cross-Lingual Disease Alignment}
\label{sec:cldarf}

For each feature $f$, COPD and control speakers are compared independently
within each language using the Mann--Whitney $U$ test ~\cite{mann1947test}. The signed
rank-biserial effect size is

\begin{equation}
r_{\ell}(f)=
\frac{2U_{\ell}(f)}{n_{\ell}^{C}n_{\ell}^{H}}-1,
\end{equation}

where positive and negative values indicate larger feature values in COPD
and controls, respectively. Benjamini--Hochberg correction ~\cite{benjamini1995controlling} is applied to
the per-language tests, while feature selection is based on signed
effect-size consistency. Disease alignment is quantified using the Language Invariance Score (LIS),

\begin{equation}
\small
\operatorname{LIS}(f)=
\begin{cases}
\min\!\left(|r_E(f)|,|r_B(f)|\right),
& \operatorname{sgn}(r_E)=\operatorname{sgn}(r_B), \\[2pt]
-\min\!\left(|r_E(f)|,|r_B(f)|\right),
& \operatorname{sgn}(r_E)\neq\operatorname{sgn}(r_B).
\end{cases}
\label{eq:lis}
\end{equation}

Positive LIS denotes preserved disease direction, while negative LIS
indicates sign reversal. The minimum absolute effect constrains alignment
by the weaker language-specific association. For threshold $\tau$,

\begin{equation}
\mathcal{F}_{\tau}
=
\{f\in\mathcal{F}:\operatorname{LIS}(f)>\tau\},
\end{equation}

defines the disease-aligned representation, while
$\operatorname{LIS}(f)<0$ defines the sign-reversing negative-control
subset.

\subsection{Evaluation Protocol}

We first evaluate within-language COPD discrimination in the Bangla cohort on the full 272-feature representation using repeated (10×) speaker-level stratified five-fold cross-validation, with imputation and standardization fitted within each training fold, and assess significance with a 200-permutation label test. We then assess CL-DAF under bidirectional English$\rightarrow$Bangla and Bangla$\rightarrow$English
 transfer. For each direction, models
are trained exclusively on the source-language cohort and evaluated
on the target-language cohort. We compare the full representation,
the sign-consistent subset, the LIS-selected disease-aligned
representation, and the sign-reversing subset under a common
classification protocol. To minimize confounding from model
complexity, all comparisons use a fixed class-balanced
L2-regularized logistic regression ($C=0.1$). Median imputation and
feature standardization are estimated from the source data and
applied unchanged to the target domain. Performance is assessed
using AUC, balanced accuracy (BAcc), sensitivity (Sen), and specificity (Spe), with
95\% confidence intervals obtained from 1,000 speaker-level
bootstrap resamples. For target-held-out validation, LIS is recomputed using the complete
source cohort and four of five target folds, while the remaining
target fold is excluded from both feature selection and classifier
training.
\vspace{-0.1in}
\section{Results}
\subsection{Within-Language COPD Detection in Bangla}

Before examining cross-lingual transfer, we first assessed whether spontaneous Bangla speech independently contains discriminative COPD-related information. As summarized in Table~\ref{tab:transfer_results}, repeated speaker-level stratified five-fold cross-validation with logistic regression achieved an AUC of $0.849 \pm 0.026$ (95\% CI: $0.755$--$0.941$), balanced accuracy of $0.765$, sensitivity of $0.704$, and specificity of $0.827$. A label-permutation test yielded a null AUC of $0.509 \pm 0.090$ ($p<0.01$), indicating performance significantly above chance. These results establish a measurable COPD-related signal in spontaneous Bangla speech and motivate the subsequent analysis of its stability under language shift.

\begin{table}[hbt]
\centering
\vspace{-0.2in}
\caption{Disease-Aligned and Sign-Reversing Acoustic Features.}
\label{tab:feature_stats}

\scriptsize
\setlength{\tabcolsep}{2.3pt}
\renewcommand{\arraystretch}{1.05}

\begin{tabular}{@{}lrrrrr@{}}
\toprule
Feature &
$r_{\mathrm{BN}}$ &
$q_{\mathrm{BN}}$ &
$r_{\mathrm{EN}}$ &
$q_{\mathrm{EN}}$ &
LIS \\
\midrule

\multicolumn{6}{@{}l}{\textbf{CL-DAF}} \\

MFCC8 $(\sigma$--$\mu)$
& -0.339 & 0.110
& -0.341 & \textbf{0.002}$^{**}$
& \textbf{0.339} \\

$\Delta$MFCC8 $(\sigma$--$\mu)$
& -0.479 & \textbf{0.040}$^{*}$
& -0.327 & \textbf{0.003}$^{**}$
& \textbf{0.327} \\

Spectral flux $(\mu$--$\mu)$
& -0.462 & \textbf{0.042}$^{*}$
& -0.291 & \textbf{0.010}$^{**}$
& \textbf{0.291} \\

$\Delta$MFCC4 $(\sigma$--$\mu)$
& -0.385 & 0.081
& -0.268 & \textbf{0.018}$^{*}$
& \textbf{0.268} \\

$\Delta$MFCC4 $(\mu$--$\sigma)$
& -0.250 & 0.289
& -0.253 & \textbf{0.029}$^{*}$
& \textbf{0.250} \\

\midrule
\multicolumn{6}{@{}l}{\textbf{Sign-reversing}} \\

Mel-band 5 energy $(\mu)$
& 0.298 & 0.156
& -0.240 & \textbf{0.036}$^{*}$
& -0.240 \\

Mel-band 5 energy $(\sigma)$
& 0.270 & 0.260
& -0.230 & \textbf{0.041}$^{*}$
& -0.230 \\

$\Delta$MFCC3 $(\sigma$--$\sigma)$
& 0.223 & 0.353
& -0.240 & \textbf{0.036}$^{*}$
& -0.223 \\

MFCC8 (median--$\sigma$)
& 0.256 & 0.272
& -0.213 & 0.058
& -0.213 \\

Mel-band 4 energy $(\mu)$
& 0.359 & 0.093
& -0.198 & 0.080
& -0.198 \\

\bottomrule
\end{tabular}

\vspace{1pt}
\begin{minipage}{\columnwidth}
\tiny
$r_{\mathrm{BN}}$ and $r_{\mathrm{EN}}$ denote signed
rank-biserial COPD--control effect sizes; $q$-values are
Benjamini--Hochberg FDR corrected.
$^{*}q<0.05$, $^{**}q<0.01$, $^{***}q<0.001$.
For notation $(a$--$b)$, $a$ denotes the within-segment
statistic and $b$ the speaker-level aggregation; the dash
denotes sequential aggregation rather than subtraction.
Positive LIS denotes consistent disease direction, whereas
negative LIS denotes sign reversal.
\end{minipage}
\vspace{-0.2in}
\end{table}

\subsection{Cross-Lingual Disease-Effect Consistency}
\label{sec:effect_consistency}

\begin{figure}[h]
    \centering
    \vspace{-0.1in}
    \includegraphics[width=0.8\columnwidth]{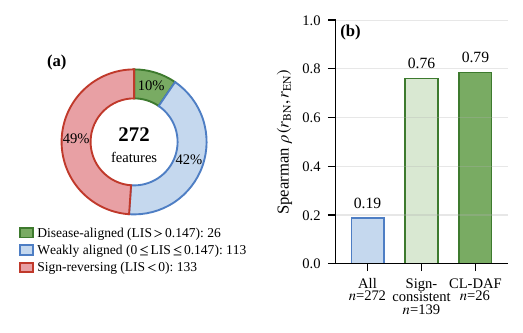}
    \vspace{-0.2in}
    \caption{ Cross-lingual disease-effect consistency. (a) Partition of the 272 features by LIS. (b)  Spearman correlation between Bangla and English
    COPD--control effect sizes for the complete representation,
    sign-consistent subset, and CL-DAF-selected subset.}
    \vspace{-0.2in}
    \label{fig:lis_stats}
\end{figure}

Cross-lingual analysis revealed substantial heterogeneity in
disease-associated feature behavior. As shown in
Fig.~\ref{fig:lis_stats}(a), of the 272 shared descriptors,
133 (48.9\%) exhibited opposite COPD--control effect directions
between Bangla and English, while 139 were sign-consistent.
Among the latter, 26 satisfied the primary disease-alignment
criterion ($\mathrm{LIS}>0.147$), while 113 exhibited weaker
but directionally consistent effects. This partition was reflected in cross-lingual effect-size
agreement [Fig.~\ref{fig:lis_stats}(b)]. Agreement across the
complete representation was weak but significant
(Spearman $\rho=0.189$, $p<0.01$), but increased markedly
among sign-consistent features ($\rho=0.761$, $p<0.001$).
The 26 CL-DAF-selected descriptors exhibited the strongest
agreement ($\rho=0.785$, $p<0.001$), indicating that
disease-aligned feature selection isolates a substantially more
consistent cross-lingual acoustic subspace.

Table~\ref{tab:feature_stats} summarizes representative
disease-aligned and sign-reversing descriptors. Among the
CL-DAF features, $\Delta$MFCC8 variability and spectral flux remained significant after Benjamini–Hochberg correction in both languages ($q<0.05$),
with concordant negative COPD effects. Other highly ranked
features preserved effect direction but reached FDR significance
in only one domain, consistent with the lower statistical power
of the smaller Bangla cohort. In contrast, prominent
sign-reversing descriptors exhibited opposing effects across
languages despite several showing significant within-language
associations, demonstrating that within-domain statistical
relevance alone does not guarantee cross-lingual disease
alignment.

\begin{table*}[t]
\centering
\caption{Cross-Lingual and Within-Language COPD Classification Performance.}
\label{tab:transfer_results}

\setlength{\tabcolsep}{3.2pt}
\renewcommand{\arraystretch}{1.08}

\begin{tabular}{lccccccccc}
\toprule
& & \multicolumn{4}{c}{\textbf{English$\rightarrow$Bangla}}
& \multicolumn{4}{c}{\textbf{Bangla$\rightarrow$English}} \\
\cmidrule(lr){3-6}
\cmidrule(lr){7-10}

Representation & $N_f$
& AUC [95\% CI] & BAcc & Sen & Spe
& AUC [95\% CI] & BAcc & Sen & Spe \\
\midrule

All Features
& 272
& 0.663 [0.531, 0.792]
& 0.606 & 0.885 & 0.327
& 0.488 [0.409, 0.569]
& 0.493 & 0.814 & 0.172 \\

Sign-consistent
& 139
& 0.779 [0.663, 0.886]
& 0.714 & 0.692 & 0.735
& 0.645 [0.567, 0.721]
& 0.595 & 0.735 & 0.455 \\

Sign-reversing
& 133
& 0.268 [0.159, 0.393]
& 0.371 & 0.538 & 0.204
& 0.341 [0.270, 0.417]
& 0.428 & 0.755 & 0.101 \\

\textbf{CL-DAF}
& \textbf{26}
& \textbf{0.825 [0.709, 0.930]}
& 0.751 & 0.808 & 0.694
& \textbf{0.722 [0.652, 0.791]}
& 0.683 & 0.598 & 0.768 \\

\midrule
\multicolumn{10}{l}{
\textbf{Bangla within-language (272 features): AUC = 0.849 [0.755, 0.941], BAcc = 0.765, Sen = 0.704, Spe = 0.827}
} \\

\bottomrule
\end{tabular}
\end{table*}

\subsection{Cross-Lingual Transfer Performance}
\begin{figure}[t]
    \centering
    \includegraphics[width=\columnwidth]{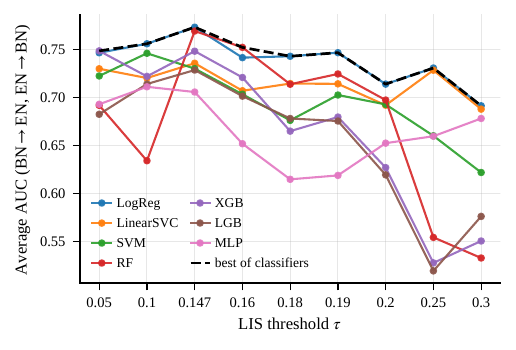}
    \vspace{-0.35in}
    \caption{ Cross-lingual AUC across LIS thresholds and classifiers.}
    \label{fig:threshold_ablation}
    \vspace{-0.2in}
\end{figure}

Despite the strong within-language COPD discrimination observed in
Bangla, cross-lingual transfer using the complete acoustic
representation remained limited. We therefore examined whether
increased disease-effect consistency translated into improved
cross-lingual classification. Table~\ref{tab:transfer_results}
compares the complete representation, sign-consistent and
sign-reversing subsets, and the CL-DAF representation at the
primary threshold $\tau=0.147$. All comparisons use the same fixed
L2-regularized logistic regression. Using all 272 descriptors yielded limited and asymmetric transfer, with AUCs of 0.663 for English-to-Bangla and 0.488 for Bangla-to-English. Restricting the representation to sign-consistent features increased AUCs to 0.779 and 0.645, respectively, whereas the sign-reversing subset performed substantially below chance, reaching 0.268 and 0.341. CL-DAF achieved the strongest bidirectional transfer while retaining only 26 features, with AUCs of 0.825 for English-to-Bangla and 0.722 for Bangla-to-English. The corresponding balanced accuracy, sensitivity, and specificity are reported in Table ~\ref{tab:transfer_results}.
Together with the effect-size analysis in Fig.~\ref{fig:lis_stats}, these results indicate that preserving the direction of disease-associated acoustic changes is important for cross-lingual transfer. To assess the robustness of this effect, we varied the LIS threshold and repeated the analysis across multiple classifier families. As shown in Fig.~\ref{fig:threshold_ablation}, disease-aligned representations maintained strong performance over a broad threshold range, particularly $\tau=0.10$--$0.19$. Similar gains across logistic regression, linear SVM, and other classifiers further suggest that the benefit of CL-DAF is primarily representation-driven rather than classifier-specific.

\subsection{Target-Held-Out Feature-Selection Validation}
Because pooled LIS uses labeled observations from both
languages to identify disease-aligned features, we additionally
performed target-held-out feature-selection validation. The
target cohort was partitioned into five folds; for each fold,
LIS was recomputed using the complete source cohort and only
the remaining target folds. The held-out target speakers
therefore contributed neither to feature selection nor classifier
training, while the logistic-regression classifier remained
trained exclusively on the source language.

As shown in Table~\ref{tab:nested_validation}, across repeated target-held-out partitions, CL-DAF retained
substantial transfer performance, achieving AUC
$0.785\pm0.015$ for EN$\rightarrow$BN and 
$0.630\pm0.022$ for BN$\rightarrow$EN . In contrast,
selecting features from source-language effect size alone
yielded weaker and asymmetric transfer (0.666 and 0.544).
The reduction relative to pooled selection (0.825 and 0.722)
quantifies the optimism introduced when the complete target
cohort participates in disease-alignment estimation, while the
target-held-out results demonstrate generalization to unseen
target speakers.
\begin{table}[hbt]
\centering
\caption{Cross-Lingual Feature-Selection Performance.}
\label{tab:nested_validation}

\scriptsize
\setlength{\tabcolsep}{3.2pt}
\renewcommand{\arraystretch}{1.07}

\begin{tabular}{lccc}
\toprule
Selection protocol & $N_f$ &
EN$\rightarrow$BN (AUC) & BN$\rightarrow$EN (AUC) \\
\midrule

Source-only
& 94 / 148
& 0.666 [0.536, 0.796]
& 0.544 [0.467, 0.624] \\

\textbf{Target-held-out LIS}
& 27.5 / 29.2
& \textbf{0.785$\pm$0.015}
& \textbf{0.630$\pm$0.022} \\

Pooled LIS
& 26
& 0.825 [0.709, 0.930]
& 0.722 [0.652, 0.791] \\

\bottomrule
\end{tabular}

\vspace{1pt}
\begin{minipage}{\columnwidth}
\tiny
For direction-dependent methods, $N_f$ is reported as
EN$\rightarrow$BN / BN$\rightarrow$EN.
Target-held-out results are mean$\pm$SD across repeated
5-fold target partitions; no held-out target speaker contributes
to LIS computation or classifier training.
\end{minipage}
\vspace{-0.2in}
\end{table}
\vspace{-0.1in}
\section{Conclusion}
This study demonstrates that spontaneous Bangla speech contains discriminative COPD-related information, extending spontaneous-speech respiratory assessment to a previously unexplored linguistic setting. CL-DAF further provides an interpretable framework for identifying acoustic features whose disease effects remain stable across languages while separating features that can hinder transfer. These findings support a broader shift from language-specific clinical speech models toward multilingual representations that prioritize pathological information over language- and phonetic-dependent variation. While the present study validates this principle across Bangla and English, future work will examine its consistency in additional languages and use these disease-aligned representations to guide the development of unified multilingual models requiring progressively less target-language supervision.
\vspace{-0.1in}
\section{Acknowledgments}
The authors gratefully acknowledge Sylhet MAG Osmani Medical College Hospital for supporting clinical data collection, and Line Reflection Ltd. for supporting this study. The authors have no relevant financial or non-financial interests to disclose.
\section{COMPLIANCE WITH ETHICAL STANDARDS}
The Bangla data collection protocol was reviewed and approved by the Institutional Review Board of the University of Maryland (IRB-2323776-2). All participants provided informed consent prior to recording.

\bibliographystyle{IEEEbib}
\bibliography{strings,refs}

\begin{thebibliography}{10}

\bibitem{fuchs2021respiratory}
Susanne Fuchs and Am{\'e}lie Rochet-Capellan,
\newblock ``The respiratory foundations of spoken language,''
\newblock {\em Annual Review of Linguistics}, vol. 7, no. 1, pp. 13--30, 2021.

\bibitem{mayr2025assessing}
Wolfgang Mayr, Andreas Triantafyllopoulos, Anton Batliner, Bj{\"o}rn~W
  Schuller, and Thomas~M Berghaus,
\newblock ``Assessing the clinical and functional status of {COPD} patients
  using speech analysis during and after exacerbation,''
\newblock {\em International Journal of Chronic Obstructive Pulmonary Disease},
  pp. 137--147, 2025.

\bibitem{nallanthighal2022detection}
Venkata~Srikanth Nallanthighal, Aki H{\"a}rm{\"a}, and Helmer Strik,
\newblock ``Detection of {COPD} exacerbation from speech: comparison of
  acoustic features and deep learning based speech breathing models,''
\newblock in {\em ICASSP 2022-2022 IEEE International Conference on Acoustics,
  Speech and Signal Processing (ICASSP)}. IEEE, 2022, pp. 9097--9101.

\bibitem{who2026copd}
{World Health Organization},
\newblock ``Chronic obstructive pulmonary disease {(COPD)},'' June 2026,
\newblock [Online]. Available:
  \url{https://www.who.int/news-room/fact-sheets/detail/chronic-obstructive-pulmonary-disease-(copd)}.
  Accessed: Sep. 1, 2026.

\bibitem{idrisoglu2024copdvd}
Alper Idrisoglu, Ana~Luiza Dallora, Abbas Cheddad, Peter Anderberg, Andreas
  Jakobsson, and Johan~Sanmartin Berglund,
\newblock ``{COPDVD}: automated classification of chronic obstructive pulmonary
  disease on a new collected and evaluated voice dataset,''
\newblock {\em Artificial intelligence in medicine}, vol. 156, pp. 102953,
  2024.

\bibitem{triantafyllopoulos2024sustained}
Andreas Triantafyllopoulos, Anton Batliner, Wolfgang Mayr, Markus Fendler,
  Florian Pokorny, Maurice Gerczuk, Shahin Amiriparian, Thomas Berghaus, and
  Bj{\"o}rn Schuller,
\newblock ``Sustained vowels for pre-vs post-treatment {COPD} classification,''
\newblock {\em arXiv preprint arXiv:2406.06355}, 2024.

\bibitem{haider2019respiratory}
Nishi~Shahnaj Haider, Bikesh~Kumar Singh, R~Periyasamy, and Ajoy~K Behera,
\newblock ``Respiratory sound based classification of chronic obstructive
  pulmonary disease: a risk stratification approach in machine learning
  paradigm,''
\newblock {\em Journal of medical systems}, vol. 43, no. 8, pp. 255, 2019.

\bibitem{khanom2026spirophonia}
Roksana Khanom, Shafia Supty, Nirupam Roy, and Ashok Agrawala,
\newblock ``Spirophonia: Non-invasive respiratory health assessment from
  spontaneous speech,''
\newblock in {\em Proc. INTERSPEECH 2026}, Sydney, Australia, 2026,
\newblock to appear, arXiv:2609.17350.

\bibitem{bhalla2025phoneme}
Sejal Bhalla, Tien Han, Andrea Gershon, Robert Wu, Eyal de~Lara, and Alex
  Mariakakis,
\newblock ``Phoneme-aware acoustic analysis of natural speech for lung function
  assessment,''
\newblock in {\em ICASSP 2025-2025 IEEE International Conference on Acoustics,
  Speech and Signal Processing (ICASSP)}. IEEE, 2025, pp. 1--5.

\bibitem{ethnologue2026}
David~M. Eberhard, Gary~F. Simons, and Charles~D. Fennig,
\newblock ``Ethnologue: Languages of the world,'' SIL International, Dallas,
  TX, 2026,
\newblock \url{https://www.ethnologue.com/language/ben/}, accessed Sep. 10,
  2026.

\bibitem{ud2010bengali}
Sameer ud~Dowla~Khan,
\newblock ``Bengali ({Bangladeshi} standard),''
\newblock {\em Journal of the International Phonetic Association}, pp.
  221--225, 2010.

\bibitem{alam2015prevalence}
Dewan~S Alam, Muhammad~Ah Chowdhury, Ali~T Siddiquee, Shyfuddin Ahmed, and
  John~D Clemens,
\newblock ``Prevalence and determinants of chronic obstructive pulmonary
  disease ({COPD}) in {Bangladesh},''
\newblock {\em COPD: Journal of Chronic Obstructive Pulmonary Disease}, vol.
  12, no. 6, pp. 658--667, 2015.

\bibitem{favaro2023multilingual}
Anna Favaro, Laureano Moro-Vel{\'a}zquez, Ankur Butala, Chelsie Motley, Tianyu
  Cao, Robert~David Stevens, Jes{\'u}s Villalba, and Najim Dehak,
\newblock ``Multilingual evaluation of interpretable biomarkers to represent
  language and speech patterns in {Parkinson's} disease,''
\newblock {\em Frontiers in Neurology}, vol. 14, pp. 1142642, 2023.

\bibitem{luz2023multilingual}
Saturnino Luz, Fasih Haider, Davida Fromm, Ioulietta Lazarou, Ioannis
  Kompatsiaris, and Brian MacWhinney,
\newblock ``Multilingual {Alzheimer’s} dementia recognition through
  spontaneous speech: a signal processing grand challenge,''
\newblock in {\em ICASSP 2023-2023 IEEE International Conference on Acoustics,
  Speech and Signal Processing (ICASSP)}. IEEE, 2023, pp. 1--2.

\bibitem{rudin2019stop}
Cynthia Rudin,
\newblock ``Stop explaining black box machine learning models for high stakes
  decisions and use interpretable models instead,''
\newblock {\em Nature machine intelligence}, vol. 1, no. 5, pp. 206--215, 2019.

\bibitem{bradlow2011language}
Ann~R Bradlow, Lauren Ackerman, L~Ann Burchfield, Lisa Hesterberg, Jenna Luque,
  and Kelsey Mok,
\newblock ``Language-and talker-dependent variation in global features of
  native and non-native speech,''
\newblock in {\em Proceedings of the... International Congress of Phonetic
  Sciences. International Congress of Phonetic Sciences}, 2011, p. 356.

\bibitem{wang2026cross}
Qingyi Wang and Meihong Wu,
\newblock ``Cross-lingual {Alzheimer’s} disease speech detection: Polarity
  inversion and few-shot calibration strategies,''
\newblock {\em Bioengineering}, vol. 13, no. 6, pp. 629, 2026.

\bibitem{jadoul2018introducing}
Yannick Jadoul, Bill Thompson, and Bart De~Boer,
\newblock ``Introducing parselmouth: A python interface to praat,''
\newblock {\em Journal of Phonetics}, vol. 71, pp. 1--15, 2018.

\bibitem{mcfee2015librosa}
Brian McFee, Colin Raffel, Dawen Liang, Daniel~PW Ellis, Matt McVicar, Eric
  Battenberg, Oriol Nieto, et~al.,
\newblock ``librosa: Audio and music signal analysis in python.,''
\newblock {\em SciPy}, vol. 2015, no. 18-24, pp. 7, 2015.

\bibitem{mann1947test}
Henry~B Mann and Donald~R Whitney,
\newblock ``On a test of whether one of two random variables is stochastically
  larger than the other,''
\newblock {\em The annals of mathematical statistics}, pp. 50--60, 1947.

\bibitem{benjamini1995controlling}
Yoav Benjamini and Yosef Hochberg,
\newblock ``Controlling the false discovery rate: a practical and powerful
  approach to multiple testing,''
\newblock {\em Journal of the Royal statistical society: series B
  (Methodological)}, vol. 57, no. 1, pp. 289--300, 1995.

\end{thebibliography}

\end{document}